\documentclass[12pt,reqno]{amsart}
\usepackage[english]{babel}
\usepackage{amsmath,amssymb,amsfonts,cite}

\newtheorem{theorem}{Theorem}

\numberwithin{equation}{section}

\let\p\partial

\def\phi{\varphi}
\let\bsy\boldsymbol
\let\ge\geqslant
\let\le\leqslant
\let\t\tilde

\def\be{\begin{equation}}
\def\ee{\end{equation}}
\def\ba{\begin{aligned}} 
\def\ea{\end{aligned}}

\def\n{\nonumber}

\newcounter{theo}

\newcounter{lem}

\newcounter{prop}
\newcounter{rem}
\newcommand{\rem}{\addtocounter{rem}{1}\textbf{Remark \therem.} }
\newcounter{defi}

\newcounter{examp}

\numberwithin{equation}{section}

\begin{document}

\baselineskip=6mm
\thispagestyle{empty}
\begin{center}
{\Large\bf Integrable evolution Hamiltonian equations \\[2mm] of the third order with the Hamiltonian operator $\bsy D_x$}
\end{center}

 \vskip5mm \hfill
\begin{minipage}{12.5cm}
\baselineskip=15pt
{\bf A.G.  Meshkov ${}^{1}$
 and
   V.V. Sokolov ${}^{2}$} \\ [1ex]
{\footnotesize
${}^1$ Orel University -- UNPK, Russia
 \\
${}^{2}$ Landau Institute for Theoretical Physics, Moscow, Russia }\\

\end{minipage}
\begin{center}
\begin{minipage}[c]{140mm}
\small ABSTRACT. All non-equivalent integrable evolution   equations of third order of the form $u_t=D_x\frac{\delta H}{\delta u}$  are found. 
\end{minipage}
\end{center}

\section{Introduction.}

We consider the third order integrable Hamiltonian evolution equations of the form 
\begin{equation}\label{Ham}
u_t=D_x\left(\frac{\delta H}{\delta u}  \right)=D_x\left(\frac{\p H}{\p u}-D_x\frac{\p H}{\p u_x}\right).
\end{equation}
Here $H(x,u,u_x)$ is the Hamiltonian and $D_x$ is the total $x$-derivative.  The celebrated KdV equation with $H=-\frac{1}{2}\,u_x^2+\frac13u^3$  provides the simplest example of such an equation. The function $H$ is defined up to equivalence
$H\to H+D_x f(x,u)+\lambda u,$ where  the function $f$ and the constant $\lambda$ are arbitrary.

Using the symmetry approach to integrability \cite{ss,mss}, we obtain a complete list of canonical forms for integrable Hamiltonians $H$. Our proof of the classification statement contains an algorithm which allows to bring any integrable Hamiltonian to one of the canonical forms by canonical transformations. 

\subsection{Canonical transformations.} Consider point transfromations of the form
\begin{equation}\label{ptr}
x=\phi(y,v), \qquad  u=\psi (y,v).
\end{equation}
The invertibility of the transformation is equivalent to the inequality  $\Delta =\psi_v\phi_y-\phi_v\psi_y\ne0$. 
Transformation (\ref{ptr}) is called {\it canonical} if $\Delta =\psi_v\phi_y-\phi_v\psi_y=1$.
It is easy to verify that canonical transformations preserve the form of equation (\ref{Ham}). 
The Hamiltonian of the resulting equation is given by
\begin{equation}\label{trH}
\t H(y,v,v_y)=H\left(\phi(y,v),\psi (y,v),\frac{D_y(\psi)}{D_y(\phi)}\right)D_y(\phi).
\end{equation}

{\bf Example.} Linear   transformations of the form
$$
 x=f(y),\qquad u=\frac{v}{f'}+g(y),\qquad \t H=Hf'.
$$
are canonical for arbitrary functions $f$ and $g$.

\rem If we consider only Hamiltonians that do not depend on $x$ explicitly, we still have non-trivial canonical transformations
$$ 
\begin{array}{c}
x=f(v)+ y,\quad u=v,\qquad \t H=(f'v_y+1)\,H.   \\[4mm]
x=v,\quad  u=f(v)+y, \qquad \t H=H\,v_y.   \qquad \square
\end{array}
$$

Becides (\ref{ptr}) we use the following canonical transformations of a more general form: 

1. Dilatations of the form
$$
 t=\alpha \t  t,\qquad x=\beta y,\qquad u=\gamma v, \qquad \t H =\frac{\alpha }{\beta \gamma^2}H(\beta y,\gamma v)
$$
are admissible for any $H$.
 
2. If $H$ does not depend on $x$,  then the Galilean transformation 
$$
y= x +c\,t,\qquad  v=u, \qquad   \t H=H-\frac{1}{2}\,c\,v^2;
$$
is admissible.

3. If $H=c\,xu+h(u_x)$, where $c$ is a constant, then the following transformation
$$
u\to u+c\,t,\quad H\to H-c\,xu
$$
is admissible.

\subsection{Integrability conditions.}  Necessary integrability  conditions  for equations of the form 
\begin{equation}\label{eqsc}
u_t=F(x,u, u_x, u_{xx}, u_{xxx})
\end{equation}
are given by  a series of conservation laws \cite{ss,mss}:
\begin{equation}\label{laws}
\frac{d}{dt}\rho_n=\frac{d}{dx}\theta_n,\ \ \ n=-1,0,1,\dots
\end{equation}
where $\rho_n$ are said to be the canonical densities. They can be  defined by  the following recursive formula presented here at the first time:
\begin{align}\label{reccur} 
\rho_{n+2}&=\frac{1}{3}\rho_{-1}\left(\theta_n-F_0\delta_{n,0}-F_1\rho_n-F_2D_x(\rho_n)-F_2\sum_{-1}^{n} \rho_i\rho_j\right)- \frac{1}{3}(\rho_{-1})^{-2}D_x^2(\rho_n) \n \\
&- (\rho_{-1})^{-2}\left(\frac{1}{2}D_x\sum_{-1}^{n} \rho_i\rho_j+\frac{1}{3}\sum_{0}^{n} \rho_i\rho_j\rho_k+\rho_{-1}\sum_{0}^{n+1} \rho_{i}\rho_j\right),\ \ \ n=-2,-1,0,\dots,
\end{align}
where $\rho_n=\theta_n=0$ for $n<-1$, $\,F_n={\p F}/{\p u_n},\ \rho_{-1}=F_3^{-1/3}$, and $\delta_{ik}$ is the Kronecker symbol.   
By definition,
$$
\sum_{a}^{b} \rho_{I_1}\cdots \rho_{I_k}=\sum_{\substack{I_s\ge a,\,1\le s\le k,\\ I_1+\cdots +I_k=b  }} \rho_{I_1}\cdots \rho_{I_k}.
$$
In particular,
$$
\begin{aligned}
&\sum_{-1}^{-2} \rho_i\rho_j=\rho_{-1}^2,\quad \sum_{-1}^{-1} \rho_i\rho_j=2\,\rho_{-1}\rho_0,\quad \sum_{0}^{-1} \rho_i\rho_j\rho_k=0,\\
&\sum_{0}^{0} \rho_i\rho_j\rho_k=\rho_0^3,\quad \sum_{0}^{1} \rho_i\rho_j\rho_k=3\,\rho_0^2\rho_1,\dots
\end{aligned}
$$

The first three canonical densities are given by  
$$
\begin{aligned}
&\rho_{-1}=F_3^{-1/3},\qquad \rho_0=-D_x(\ln\rho_{-1})-\frac13 F_2\rho_{-1},\\ 
&\rho_1=\frac13 \theta_{-1}\rho_{-1}-\frac13 F_1\rho_{-1}^2+F_2\rho_{-1}D_x(\rho_{-1})+\frac19 F_2^2\rho_{-1}^5+\frac13\rho_{-1}^2D_x(F_2)\\
&\qquad+\frac23 D_x(\rho_{-1}^{-2}D_x(\rho_{-1}))+\frac13 \rho_{-1}^{-3}(D_x(\rho_{-1}))^2.
\end{aligned}
$$
Notice that the flux $\theta_{-1}$ of the first canonical conservation law (\ref{laws}) is involved in the formula for $\rho_1$.

The integrability conditions lead to some partial differential equations for the right hand side $F$ of  
(\ref{eqsc}). We don't explain here how to derive these PDEs (see for example \cite{sokmesh}, where this technique is described in detailes). 

For equations (\ref{Ham}) any canonical density $\rho_n$ can be expressed in terms of the Hamiltonian $H$ and  $\theta_{-1}, \theta_0, \dots,  \theta_{n-3}, \theta_{n-2} $ . In particular, 
$$
\rho_{-1}=-\left(\frac{\p^2 H}{\p u_1^2}\right)^{-1/3}.
$$
Let us denote $\rho_{-1}=a$. Then 
$$
\frac{\p^2 H}{\p u_1^2}=-a^{-3},\qquad a=a(x,u,u_x). 
$$

The integrability conditions provide PDEs for $H$, which allow us to obtain a complete list of integrable Hamiltonians.
It is known \cite{ss} that for Hamiltonian equations all even integrabiliry conditions are trivial. Almost all the information of 
integrable Hamiltonians will be derived for the fist and third integrability conditions. 

\section{Classification statement.}

{\theorem Any non-linear equation of the form {\rm (\ref{Ham})} that has infinite hierarchy of higher symmtries 
$$
u_{\tau_k}=F_k(x,u,u_x,\dots),\ \ \ \ k=1,2,\dots
$$
is canonically equivalent to one of the following equation:
\begin{align}
&u_t=D_x\left(\frac{u_{xx}}{a^3 }-\frac{3\,a'}{2\,a^4}\,u_x^2+\frac{\p }{\p u}\frac{P(u)}{a}\right),\qquad  H= -\frac{u_x^2}{2\,a^3}+\frac{P(u)}{a}, \ \ \ \label{e-a0} \\
&\qquad \text{where}\ a=c_1u^2+c_2u+c_3,\ \n \\[4mm]
\label{e-a2-2}
&u_t=D_x\left(\frac{u_{xx}}{u^3 }-\frac{3\,u_x^2}{2\,u^4}+P(x) u^2\right),\qquad H=-\frac{u_x^2}{2\,u^3 }+\frac13 P(x) u^3,\\[3mm]
&u_t=D_x\left(\frac{u_{xx}}{{\sqrt{u_x+P(u)}^{\ 3}}}+3\,\frac{P'(u)}{\sqrt{u_x+P(u)}}-\frac{P(u) P'(u)}{(u_x+P(u))^{ 3/2}}\right),\qquad   H=4 \sqrt{u_x+P(u)}. \label{eq-c1-a1} 
\end{align}Here  $P$ is an  arbitrary polynomial of degree not greater then 4,\ $c_i$  are arbitrary constants.} $\square$

\rem Using translations $u\to u+c$, dilatations $u\to\lambda  u,\ t\to\alpha t,\, x\to \beta x,$ and the Galilean transformation, one can reduce (\ref{e-a0}) to one of the following canonical forms:  
\begin{align}\label{e-a1-1}
&u_t=D_x(u_{xx}+u^3),\qquad  H=-\frac{1}{2}\,u_x^2+\frac14u^4, \tag{\ref{e-a0}a}\\
\label{e-a1-2}
&u_t=D_x(u_{xx}+u^2),\qquad  H=-\frac{1}{2}\,u_x^2+\frac13u^3,  \tag{\ref{e-a0}b} \\
\label{e-a2-1}
&u_t=D_x\left(\frac{u_{xx}}{u^3 }-\frac{3\,u_x^2}{2\,u^4}+c_1u^2+\frac{c_2}{u^2}\right),\qquad 
H=-\frac{u_x^2}{2\,u^3 }+\frac13c_1u^3-c_2u^{-1},    \tag{\ref{e-a0}c}  \\
\label{e-a3}
&u_t=D_x\left(\frac{u_{xx}}{a^3 }-3\frac{uu_x^2}{a^4}+c_1\frac{c-u^2}{a^2}-2\,c_2\frac{u}{a^2}\right),\quad
H=-\frac{u_x^2}{2\,a^3 }+\frac{c_1u+c_2}{a },  \tag{\ref{e-a0}d}  
\end{align}
where  $ a=u^2+c. \ \square $

{\bf Proof of Theorem 1.}  It follows from the first integrability condition ($n=-1$ in (\ref{laws})) that 
\begin{equation}\label{e0}
\frac{d}{dx}\left(a^3\frac{\p^2 a}{\p u_x^2}\right)=0.
\end{equation} 
The  solution of (\ref{e0}) is given by  
\begin{equation}\label{eq1}
a=\sqrt{a_1u_x^2+a_2u_x+a_3},
\end{equation}
where $a_i=a_i(x,u)$ and
\begin{equation}\label{eq2}
a_2^2-4\,a_1a_3=const.
\end{equation}

Under canonical tansformations the function $a$ transforms as follows
$$
\t a=\sqrt{\t a_1v_y^2+\t a_2v_y+\t a_3}\,,\qquad\text{where}\ \ \t a_1=a_1\phi_x^2-a_2\phi_x\phi_u+a_3\phi_u^2.
$$
Hence we can reduce $a_1$ to zero by an appropriate canonical transformation.  Taking into account (\ref{eq2}), we see that it suffices to consider the following  two cases: 
$({\bf A}) \ a=a(x,u)$ \  {and} \ $({\bf B})  \ a=\sqrt{ u_x+q(x,u)}\quad$ (if  $\t a_1=0$ then $\t a_2$ is a constant, which can be brought  to 1 by a dilatation).  \\

\subsection{\bf Case A}  In this case the Hamiltonian is given by  
\begin{equation}\label{h-a}
H=h(x,u)-\frac{u_x^2}{2\,a^3},\qquad a=a(x,u).
\end{equation}
The first integrability condition implies the following diffential equations 
$$
\frac{\p^3a}{\p u^3}=0,\qquad \frac{\p}{\p x}\left(\frac{\p^2 a^2}{\p u^2}-3\left(\frac{\p a}{\p u}\right)^2\right)=0,
$$ 
therefore
\begin{equation}
a=s_1 u^2+s_2 u+s_3,\qquad   s_i=s_i(x), \label{a-A}
\end{equation} 
and 
\begin{equation}\label{eq3}
 \frac{d}{dx}(s_2^2-4\,s_1 s_3)=0.
\end{equation}
It is easy to verify that we can reduce the functions $s_1(x), s_2(x), s_3(x)$ to constants  by a canonical tranformation of the form  
$$
 x=f(y),\qquad u=\frac{v}{f'}+g(y).
$$
So we obtain
 \begin{align}\label{h-a}
H=h(x,u)-\frac{u_x^2}{2\,a^3},\qquad a=c_1u^2+c_2u+c_3.
\end{align}

The third integrability condition implies  
$$
a\,\frac{\p^5 h}{\p u^5}+5\,a'\,\frac{\p^4 h}{\p u^4}+10\,a''\frac{\p^3 h}{\p u^3}=0.
$$
Substituting    $g/a$ for $h$, we get $g^{(5)}=0$ and therefore 
\begin{equation}\label{Hgen}
H=\frac{r_1u^4+r_2u^3+r_3u^2+r_4u+r_5}{c_1u^2+c_2u+c_3}-\frac{u_x^2}{2\,(c_1u^2+c_2u+c_3)^3 },\qquad r_i=r_i(x).
\end{equation}

To determine the $x$-dependence of the functions $r_i(x)$ we consider several subcases.

{\bf Subcase A.1.} Let $c_1=c_2=0$. Then the Hamiltonian is  equivalent to  
$$
H=-\frac{1}{2}\,u_x^2+\frac14q_1u^4+\frac13q_2u^3+\frac12q_3u^2+q_4u,
$$
where $q_i=q_i(x)$.  The third integrability condition is equivalent to relations 
\begin{equation}\label{eq-a1}
q_1=k_1,\qquad 2\,q_2q_2'=3\,k_1 q_3',\qquad 2\,q_2'''+2\,q_2'q_3=4\,k_1q_4',
\end{equation}
where $k_1$ is a constant.

If $k_1\ne0$, then we normalize it to 1 by $u\to u\, k_1^{-1/2}$ and reduce $q_2$ to zero by $u\to u-q_2(x)/3$.  It follows from (\ref{eq-a1}) that now we have  $q_3'=q_4'=0$. Thus the Hamiltonian is equivalent to  (\ref{e-a1-1}) and we arrive at the modified KdV equation.

If $k_1=0$, then (\ref{eq-a1}) implies $q_2=k_2$, where $k_2$ is a constant. Suppose that $k_2\ne0$. Then we normilize $k_2$ to 1 and reduce  $q_3$ to zero by $u\to u-q_3(x)/2$. It follows from the fifth integrability condition that $q_4'=0$ and we obtain the KdV equation (\ref{e-a1-2}).

At last, if  $k_1=k_2=0$, then corresponding equation becomes linear:
\begin{equation}\label{e-a1-3}
u_t=D_x(u_{xx}+f(x)u+g(x)).
\end{equation}
where $f$ and $g$ are arbitrary functions.

{\bf Subcase A.2.}  Suppose $c_1=0, c_2\ne 0$. Then the Hamiltonian is  equivalent to  

$$
H=-\frac{u_x^2}{2\,u^3 }+\frac13q_1u^3+\frac12q_2u^2-q_3u^{-1}+q_4u,
$$
where  $q_1,q_2,q_3$ and $q_4$ are functions of $x$. It follows from the third integrability condition that  
\begin{equation}\label{eq-a2}
q_1'q_2-2\,q_1q_2'=0,\qquad q_2'''-q_2q_3'-2\,q_2'q_3=0,\qquad q_4'=0.
\end{equation}
Since $q_4$ is a constant, without loss of generality we put $q_4=0$. 

Under canonical transformation 
$$y=f(x),\qquad v=u/f'$$ 
the Hamiltonian transforms as follows: 
$$
\t H=-\frac{v_y^2}{2\,v^3 }+\frac13q_1{f'}^2v^3+\frac12q_2f'v^2-(v\,{f'}^4)^{-1}\left(f'f'''-\frac32{f''}^2+q_3{f'}^2\right).
$$
 Consider the following cases:
\textbf{(\textit{a})} $q_2\ne0$ and  \textbf{(\textit{b})} $q_2=0.$ 

In the case \textbf{(\textit{a})} taking $\int 1/q_2\, dx$ for $f$, we bring $q_2$ to 1.  Then it follows from (\ref{eq-a2}) that 
 $q_1'=q_3'=0$ and we get an equation equivalent to (\ref{e-a2-1}). 

In the case  \textbf{(\textit{b})} taking for $f$ any nonconstant solution of equation  $f'f'''-\frac32{f''}^2+q_3{f'}^2$, we reduce $q_3$ to zero. Then the fifth integrability 
condition implies  $q_1^{(5)}=0$ and we arrive at equation (\ref{e-a2-2}).

\rem If we consider the normalization $q_3=0$ instead of $q_2=1$  in the case \textbf{(\textit{a})}, then
we get
$$
u_t=D_x\left(\frac{u_{xx}}{u^3 }-\frac{3\,u_x^2}{2\,u^4}+c\,q_2^2(x) u^2+q_2(x) u\right),\qquad H=-\frac{u_x^2}{2\,u^3 }+\frac13 c\,q_2^2(x)u^3+\frac12 q_2(x) u^2,\quad q_2'''=0 .
$$
We have chosen the canonical form (\ref{e-a2-1}) since the corresponding Hamiltonian does not depend on $x$ explicitly. $\square$

\rem  If we consider  the normalization $q_1=1$ instead of $q_3=0$  in the case $(\bsy b),$ then we obtain
$$
u_t=D_x\left(\frac{u_{xx}}{u^3 }-\frac{3\,u_x^2}{2\,u^4}+u^2+\frac32\frac{\wp(x)}{u^2}\right),
$$
where  ${\wp'}^2=4\,\wp^3-g_2\,\wp-g_3,\ {\wp'}\ne0$. It is another canonical form for equation  (\ref{e-a2-2}). $\square$

{\bf Subcase A.3.}    Suppose $c_1\ne 0$ in (\ref{Hgen}).  Using the dilatation $u\to u c_1^{-1/2},$ we normalize $c_1$ to 1. Then the translation $u\to u-c_2/2$ reduces $a$ to the form $a=u^2+c$. In this case the first integrability condition yields $\p H/\p x=0$. 
Therefore all the functions $r_i$ in (\ref{Hgen}) are constants and the equation is equivalent to (\ref{e-a3}).

\subsection{\bf Case B}  Consider Hamiltonians of the form
$$
H=h(x,u)+4\,a,\qquad a=\sqrt{u_x+q(x,u)}.
$$
It follows from the first integrability condition that 
\begin{align}
&\frac{\p^3 h}{\p u^3}=0,  \label{eqb1}\\ 
&\frac{\p^3h}{\p x^2\p u}-2\,q\,\frac{\p^3h}{\p x\p u^2}+\frac{\p q}{\p u}\,\frac{\p^2h}{\p x\p u}-\frac{\p q}{\p x}\,\frac{\p^2h}{\p u^2}=0.\label{eqb2}
\end{align}
The third integrability condition implies one more simple PDE: $\ \p^5 q/\p u^5=0$. Solving this equation and (\ref{eqb1}) we find that  
\begin{equation}\label{eqc}
q=q_1u^4+q_2u^3+q_3u^2+q_4u+q_5,\qquad h=\frac12\,h_1u^2+h_2u+h_3, 
\end{equation}
where $q_i=q_i(x),\  h_i=h_i(x)$. Substituting $q$ and $h$ in (\ref{eqb2}) we obtain the following system:
\begin{equation}\label{eqs-c1}
\begin{aligned}
&2\,q_1h_1'-q_1'h_1=0,\qquad q_2h_1'-q_2'h_1+4q_1h_2=0, \qquad q_3'h_1-3\,q_2h_2'=0,\\[2mm]
& h_1''+2\,q_3h_2'-q_4h_1'-q_4'h_1=0,\qquad  h_2''-2\,q_5h_1'+q_4h_2'-q_5'h_1=0.
\end{aligned}
\end{equation}

The canonical transformation  
$$
y=\phi (x),\qquad v=\frac{u}{\phi'}+\psi(x), 
$$
changes the Hamiltonian as follows:
$$
\t H=\frac12\,\t h_1v^2+\t h_2v+\t h_3+4\,\sqrt{v_y+\t q}\,,
$$
where
\begin{equation}\label{tr-c1}
\begin{aligned}
&\t h_1=\phi ' h_1,\qquad \t h_2=h_2-\phi' \psi h_1,\qquad \t q=Q+(\phi ')^{-2}\big(\phi ''v-(\psi \phi ')'\big),\\[2mm] 
&Q=q_1(\phi ')^2(v-\psi)^4+q_2\phi '(v-\psi)^3+q_3(v-\psi)^2+q_4(\phi ')^{-1}(v-\psi)+q_5(\phi ')^{-2}.
\end{aligned}
\end{equation}

If $h_1\ne0$ then we put $\phi'=1/h_1$ and $\psi=h_2$ to get $h_1=1$ and $h_2=0.$ Now it follows from (\ref{eqs-c1}) that $\p q_i/\p x=0$. Since $H$ does not depend on $x$ we  remove the term $\frac12u^2$ in $H$ by the Galilean transformation and obtain equation  (\ref{eq-c1-a1}).

If $h_1=0$ but $h_2'\ne0$ then equations  (\ref{eqs-c1}) lead to $q_1=q_2=q_3=0,\ h_2''+q_4h_2'=0.$ It follows from the formula
$$
\t q=\frac{v}{{\phi '}^2}(\phi''+q_4\phi')+\frac{1}{{\phi'}^2}\big(q_5-(\psi \phi')'\big)
$$ 
that there exist  $\phi $ and $\psi$ such that $\t q=0$. In this case $h_2=cx$, where $c$ is a constant. So, we obtain
\begin{align*}
H=c\,x\,u+4\,\sqrt{u_x}\,, \qquad u_t=D_x\left(\frac{u_{xx}}{u_x^{3/2}}+c\,x\right).
\end{align*} 
Using the transformation $u\to u+ct,$ we bring $c$ to zero and arrive at a particular case of equation (\ref{eq-c1-a1}).
 
If $h_1=h_2'=0$ then without loss of generality we put
$h_2=h_3=0$.  Thus, we have shown that in all cases the functions $h_1,h_2,h_3$ can be reduced to zeros.

Now we normalize the polynomial $q$ то prove that all coefficients of $q$ can be reduced to constants. If $q_1\ne0$ we use the normalization   $q_1=1,\, q_2=0$. If $q_1=0,\,q_2\ne0$ then
we normailize $q_2$ and  $q_3$ by 1 and 0 correspondingly. In the case $q_1=q_2=0$  we may put $q_4=q_5=0$. In each of these cases the third integrability condition gives rise to $q_i'=0$ for all
remaining coefficients of $q$. The corresponding equations can be obtained from  (\ref{eq-c1-a1}) by translations $u\to u+c$ and
dilatations $u\to\lambda  u,\ t\to\alpha t,\, x\to \beta x$. \qquad $\square$

\subsection{Integrability of equations (\ref{e-a0})--(\ref{eq-c1-a1})} Equations (\ref{e-a1-1}) and (\ref{e-a1-2}) are known to be integrable by the inverse scattering method. 
Equations (\ref{e-a0}c), (\ref{e-a0}d)  and (\ref{eq-c1-a1}) can be reduced to known integrable equations of the form \cite{svsok}
\begin{equation}\label{consep}
v_t=v_{yyy}+G(x,v,v_y,v_{yy})
\end{equation}
by the standard reciprocal transformation (see \cite{mss}, section 1.4)
\begin{equation}\label{reci}
dy=\rho_{-1}dx+\theta_{-1}dt, \qquad v(t,y)=u(t,x), 
\end{equation}
where $\rho_{-1}$ is the first canonical density and $\theta_{-1}$ is the correspondent flux.  Notice that this transformation is always applicable if $\rho_{-1}$ depends on $u$ only and the r.h.s. of the equation does not depend on $x$. Sometimes (\ref{reci}) can be applied in the case when $\rho_{-1}$ depends on $u$ and $u_x$. That is the case for equation (\ref{eq-c1-a1}).
 
To reduce the equation for $v$ to an usual form some additional point transformation $v=f(w)$ can be needed.
For equation (\ref{e-a2-1}) we have $\rho_{-1}=u.$ Taking $v=e^w$, we obtain  
$$
w_t=w_{yyy}-\frac{1}{2}w_y^3+w_y\left(c_1e^{2w}-3\,c_2e^{-2w}\right).
$$
This equation was found by F.Calogero and A. Degasperis and independently by A. Fokas. 

In the case of equation (\ref{e-a3}) we have $\rho_{-1}=u^2+c$. If $c\ne0$ we put $c=-k^2/4,\ v=\frac{k}{2}\tanh(w/2).$ As a result we get the same (up to the Galilean transformation) 
Calogero--Degasperis equation
$$
w_t=w_{yyy}-\frac{1}{2}w_y^3+w_y\left(\t c_1e^{2w}+\t c_2e^{-2w}\right)-c_3w_y,
$$
where $\t c_1=3/2\,k^{-2}(2\,c_2+k\,c_1),\ \t c_2=3/2\,k^{-2}(2\,c_2-k\,c_1),\ c_3=6\,c_2\,k^{-2}$. In the case $c=0$ we put $v=1/w$ to obtain the mKdV equation:
$$
w_t=w_{yyy}+12\,c_2w^2w_y+6\,c_1ww_y.
$$ 

Equation (\ref{eq-c1-a1}) is related to one more  equation found by Calogero and Degasperis:
$$
v_t=v_{yyy}-\frac{3}{8}\frac{\big(D_y(Q+v_y^2)\big)^2}{v_y(v_y+Q)}+\frac{1}{2}Q''v_y,\qquad Q=4P,
$$
by the transformation (\ref{reci}). 

Since the rigth hand side of equation (\ref{e-a2-2}) depends on $x$ we can not apply transformation  (\ref{reci}) straighforwardly. Instead we perform the substitution  $u\to u_x$ to get the potential form 
$$
u_t=\frac{u_{xxx}}{u_x^3 }-\frac{3\,u_{xx}^2}{2\,u_x^4}+P(x) u_x^2.
$$
The hodograph transformation $y=u(t,x),\ v(t,y)=x$ brings the latter equation to  the Krichever -- Novikov equation 
$$
v_t=v_{yyy}-\frac{3 v_{yy}^2}{2 v_y}-\frac{P(v)}{v_y}. 
$$

\bigskip

{\bf Acknowledgments.}
The authors would like to thank   B. Dubrovin and M. Pavlov for useful discussions. 
The research was partially supported by the RFBR grant 14-01-00751.  
VS is thankful to IHES for its support and hospitality.

\end{document}